\documentstyle[preprint,aps,prd]{revtex}
\newcommand{\be}{\begin{equation}}
\newcommand{\ee}{\end{equation}}
\newcommand{\ba}{\begin{eqnarray}}
\newcommand{\ea}{\end{eqnarray}}
\newcommand{\ft}{\tilde F_{\mu}}

\newcommand{\fmn}{F_{\mu\nu}}
\newcommand{\oh}{\displaystyle{\frac{1}{2}}}
\begin{document}
\title          {On the Field Strength Formulation of Effective
$QED_3$.}
\author{C.D. Fosco\thanks{fosco@cab.cnea.edu.ar}}
\address{ Centro At\'omico Bariloche\\
8400 Bariloche, Argentina}
\maketitle
\begin{abstract}
Halpern's field strength's formulation of gauge theories is applied
to effective $QED_3$, namely, a gauge invariant theory for an
Abelian gauge field $A_\mu$ with non-localities and self-interactions.
The resulting description in terms of the pseudovector field
$\ft = \epsilon_{\mu\nu\lambda}\partial_\nu A_\lambda$ is applied
to different examples.
\end{abstract}

\bigskip

\newpage
The search for gauge-invariant descriptions of gauge theories is a
subject with a long-standing history. A natural advantage of
gauge-invariant formulations is that, when a calculation can be
performed within such a scheme, the result is not obscured by the
unphysical features introduced in any non-explicitly gauge-invariant
setting.
In an explicitly gauge-invariant approach, the proper variables
should be first identified, and then the dynamics in terms of those
variables reconstructed. In general,
this procedure is rendered difficult because gauge-invariant variables
may be non-local~\cite{mig1,mig2}, or satisfy extremely complicated
equations.
An interesting formulation has been proposed by Halpern in the
seventies~\cite{hal1,hal2,hal3,hal4}, the so-called field-strength
formalism.  There are other interesting gauge invariant formulations. 
In the one proposed in references \cite{ha1,ha2}, a procedure to
build the Hilbert space in terms of local gauge-invariant variables
is explained.  
In this letter, we shall apply Halpern's proposal to a simple case
where the gauge-invariant description is easily constructed.  
The case we
consider is the dynamics of an Abelian gauge field in $2+1$ Euclidean
dimensions, without external matter sources. 
This does not mean that
matter
fields are absent, but rather that they could have been integrated out,
yielding a contribution to the gauge-field action that can be non-local
and
non-polynomial in general. Such kind of model has been studied in
\cite{ait}, and from a slighty different point of view 
in references \cite{mo1,mo2}, which deal with the cases of compact 
and non-compact $QED_3$.

The cases where the dynamics of a vector field
in $2+1$ dimensions is dictated by either a Maxwell or a Yang-Mills
action have already been considered by Halpern. Our study is concerned
with
a situation which is, so to speak, halfway between those cases, since
our field will be Abelian, but its action non-quadratic (and generally 
non-local).

The relevance of this kind of model comes
from the many applications $2+1$ dimensional theories have,
particularly in the realm of Condensed Matter systems~\cite{frad}.
The gauge-invariant variable in this case can be identified as
the field strength $\fmn$, or better its dual $\ft$, and one constructs
a description in terms of this pseudo-vector field. A general action
for this field will contain terms involving $\ft$ and its derivatives,
and the functional integral corresponding to it shall include a
delta-functional of the Bianchi condition $\partial \cdot {\tilde F}$.
Due to the property, particular to $3$ dimensions, of the dual of
$\fmn$ being a pseudovector, we are lead to a theory corresponding to
a self-interacting pseudovector field, which is constrained to be
transverse.

We begin by reviewing Halpern's derivation, with some small differences
due to the action not necessarily being the Maxwell one. The generating
functional
for Euclidean Green's functions of an Abelian gauge field $A_\mu$, with
a general
(possibly non-local) gauge-invariant action in $3$ dimensions is
\be
{\cal Z}(J_\mu) \;=\; \int \,{\cal D} A_\mu \, e^{- S_{inv} (A) + \int
d^3 x
J_\mu (x) A_\mu (x) }
\label{defz}
\ee
where $S_{inv} (A)$ satisfies $S_{inv} (A + \partial \omega)=S_{inv}
(A)$,
for any $\omega$ vanishing at infinity.
Of the many possible forms for $S_{inv}$ we can construct, a first
classification we make is to distinguish between parity-conserving
and parity-violating ones, since this property strongly
determines the form of the terms that can be included in $S_{inv}$.
Let us first discuss the parity-conserving case. With this assumption,
the most general form for $S_{inv}$ would be an arbitrary functional of
$\fmn$, whose terms involve contractions of different powers of
this tensor. We chose to work in terms of $\ft$, the dual of $\fmn$,
defined by $\ft = \epsilon_{\mu\nu\lambda} \partial_\nu A_\lambda$. Thus
\be
S_{inv} (A) \;=\; I(\ft)
\label{relsi}
\ee
where $I$ is an arbitrary functional. We now include into (\ref{defz})
the gauge-fixing factor corresponding to the Landau gauge ($\partial
\cdot A = 0$)
\be
{\cal Z}(J_\mu) \;=\; \int \,{\cal D} A_\mu \, \delta (\partial \cdot A)
e^{- S_{inv} (A) + \int d^3 x J_\mu (x) A_\mu (x) }
\label{gfixz}
\ee
where we have omitted the field-independent Faddeev-Popov factor
$\det (-\partial^2)$, since in this case it can be absorbed into the
normalization of the integration measure and has no effect on the
Green's functions derived from (\ref{gfixz}).
To obtain a formulation in terms of $\ft$, we introduce in (\ref{gfixz})
a `1' written as follows:
\be
1 \;=\; \int \,{\cal D} \ft \, \delta (\ft - \epsilon_{\mu\nu\lambda}
\partial_\nu A_\lambda) \, \delta (\partial \cdot {\tilde F}) \;.
\label{one}
\ee
Note the presence of a delta functional of the Bianchi identity,
which is a consistency condition for the equation
$\ft - \epsilon_{\mu\nu\lambda}\partial_\nu A_\lambda=0$, whose
solutions are relevant to the first delta-function. The meaning
of the inclusion of that factor can be made explicit by means
of the following argument: Consider the rhs of Equation
(\ref{one}), but this time writing both delta-functionals
in terms of functional Fourier transforms:
$$\int \,{\cal D} \ft \, \delta (\ft - \epsilon_{\mu\nu\lambda}
\partial_\nu A_\lambda) \, \delta (\partial \cdot {\tilde F}) $$
\be
=\, \int \,{\cal D}\ft \, {\cal D}\lambda_\mu \, {\cal D}\theta \,
\exp \left\{ i \int d^3 x [ \lambda_\mu ( \ft - \epsilon_{\mu\nu\rho}
\partial_\nu A_\rho )\,+ \, \theta \partial_\mu \ft ] \right\}
\label{expone}
\ee
where $\lambda_\mu$ and $\theta$ are Lagrange multipliers.
Integrating out $\ft$ in (\ref{expone}) yields
$$\int \,{\cal D} \ft \, \delta (\ft - \epsilon_{\mu\nu\lambda}
\partial_\nu A_\lambda) \, \delta (\partial \cdot {\tilde F}) $$
$$=\, \int \, {\cal D}\lambda \, {\cal D}\theta \,
\delta (\lambda_\mu - \partial_\mu \theta) \exp \left( -i \int d^3 x \,
\lambda_\mu \epsilon_{\mu\nu\rho} \partial_\nu A_\rho \right)$$
\be
=\, \int {\cal D}\theta \, \exp \left( -i \int d^3 x \partial_\mu
\theta \, \epsilon_{\mu\nu\rho} \partial_\nu A_\rho \right)\;=\;
\int {\cal D}\theta \, \exp \left( i \int d^3 x \theta \,
\epsilon_{\mu\nu\rho} \partial_\mu \partial_\nu A_\rho \right) \;,
\label{exinone}
\ee
where the vanishing of $\fmn$ at infinity was used on the last
line, in order to ignore the surface contribution. We conclude,
after integrating out $\theta$ in $(\ref{exinone})$ that
$$\int \,{\cal D} \ft \, \delta (\ft - \epsilon_{\mu\nu\lambda}
\partial_\nu A_\lambda) \, \delta (\partial \cdot {\tilde F}) $$
\be
=\, \delta (\epsilon_{\mu\nu\rho} \partial_\mu \partial_\nu A_\rho )\;.
\ee
Thus the `1' behaves as a constant factor when inserted into
a functional integration over $A_\mu$ fields whose second
partial derivatives commute\footnote{We are ignoring $\delta (0)$
factors.}.

After insertion of the `1', the generating functional becomes
\be
{\cal Z}(J_\mu) \;=\; \int \,{\cal D} A_\mu \, {\cal D}\ft \,
\delta (\partial \cdot A)\, \delta (\partial \cdot {\tilde F})\,
\delta (\ft - \epsilon_{\mu\nu\lambda}\partial_\nu A_\lambda)
e^{- I (\ft) + \int d^3 x J_\mu (x) A_\mu (x) } \;.
\label{gfixz}
\ee
Now we realize that, by using the two delta-functionals
$\delta (\partial \cdot A)$ and
$\delta (\ft - \epsilon_{\mu\nu\lambda}\partial_\nu A_\lambda)$,
$A_\mu$ can be written in terms of $\ft$:
\be
A_\mu \;=\; -\epsilon_{\mu\nu\lambda} \frac{1}{\partial^2}
\partial_\nu {\tilde F}_\lambda \;,
\label{fta}
\ee
and the dependence on $A_\mu$ (only from the source term) can be
completely erased by replacing it by its expression (\ref{fta})
in terms of $\ft$. The $A_\mu$ field is thus integrated out, yielding
for ${\cal Z}$ the expression:
\be
{\cal Z}(J_\mu) \;=\; \int \,{\cal D}\ft \, \delta (\partial
\cdot {\tilde F})\,
e^{- I (\ft) - \int d^3 x J_\mu \epsilon_{\mu\nu\lambda}
\partial_\nu \partial^{-2} {\tilde F}_\lambda } \;,
\label{zfin}
\ee
which contains only $\ft$ as dynamicaly variable, and may be
thought of as the generating functional for a theory describing
the dynamics of a pseudovector field $\ft$, with  the
constraint $\partial \cdot {\tilde F}=0$. We note that,
because of the form of the source term in (\ref{zfin}),
there is a simple relation between Green's functions for
$\ft$ and the ones for $A_\mu$:
$$\langle A_{\mu_1}(x_1) A_{\mu_2}(x_2)  \cdots
A_{\mu_n}(x_n)\rangle $$
\be
=\; \epsilon_{\mu_1 \nu_1
\lambda_1}\partial^{x_1}_{\nu_1}\partial_{x_1}^{-2}
\;
\epsilon_{\mu_2 \nu_2
\lambda_2}\partial^{x_2}_{\nu_2}\partial_{x_2}^{-2}
\cdots
\epsilon_{\mu_n \nu_n
\lambda_n}\partial^{x_n}_{\nu_n}\partial_{x_n}^{-2}
\langle
F_{\lambda_1}(x_1) F_{\lambda_2}(x_2) \cdots F_{\lambda_n}(x_n)
\rangle \;.
\label{relgf}
\ee

Although a naive look at (\ref{zfin}) may suggest that it is
tantamount to a gauge fixed version for some gauge-invariant
theory, this is not necessarily the case, as the general form
of the `action' $I$ for the pseudovector field is arbitrary.

We now deal with the parity-violating case. The crucial difference
with the previous discussion is that, when parity is violated,
(\ref{relsi}) is no longer valid. The reason is that now
we are allowed to include Chern-Simons like terms, which
are functions not only of $\ft$, but also of $A_\mu$, namely
\be
S_{inv} (A) \;=\; I(\ft,A) \;.
\label{relsi1}
\ee
However, an analogous procedure to the one carried out for the
parity-conserving case can be followed here, since $A_\mu$
can also be expressed in terms of $\ft$ as in (\ref{fta}).
This expression for $A$ in terms of $\ft$ is then inserted into
(\ref{relsi1}), and the generating functional for the
parity-violating case becomes:
\be
{\cal Z}(J_\mu) \;=\; \int \,{\cal D}\ft \, \delta (\partial
\cdot {\tilde F})\,
e^{- I (\ft, -\epsilon_{\mu\nu\lambda}
\partial_\nu \partial^{-2} {\tilde F}_\lambda) -
 \int d^3 x J_\mu \epsilon_{\mu\nu\lambda}
\partial_\nu \partial^{-2} {\tilde F}_\lambda } \;.
\ee
Thus, to calculate correlation functions of $\ft$,
both for the parity-conserving and parity-violating cases,
one has a generating functional corresponding to an
`action' $I$ which is a functional of $\ft$, with the
constraint $\partial \cdot {\tilde F} = 0$.

In order to do actual calculations with the theory defined
in terms of $\ft$, a  set of Feynman rules should be
defined.
It is convenient to introduce a Lagrange multiplier
field $\theta$ in order to deal with the delta-functional
$\partial \cdot {\tilde F}$,
and also to add a source term for $\theta$, since $\ft$
and $\theta$ are coupled. We add a source term for $\ft$ (not to be
confused with the source for $A_\mu$),
since the Green's functions for $A$ may be obtained by
applying (\ref{relgf}) to the $\ft$'s Green's functions.

Thus the generating functional we define is
\be
{\cal Z} \;=\; \int \,{\cal D}\ft \, {\cal D}\theta \,
\exp \left\{ - \int d^3 x [ I(\ft) - i \theta
\partial \cdot {\tilde F} \,-\,J_\mu \ft - j_\theta
\theta ] \right\}
\label{ztot}
\ee
and Euclidean correlation functions are simply obtained
by functional differentiation.
Free propagators are obtained from evaluation of the
Gaussian integral corresponding to a quadratic action,
which in the parity-conserving case becomes
\be
I(\ft) \; \equiv \; I_0(\ft) \; = \; \int d^3 x
\oh \; \ft \, D(-\partial^2) \, \ft
\label{ifree}
\ee
with $D$ a given function without real poles. It is immediate
to extract  the (momentum space) free propagators that follow
from (\ref{ztot}) with the action (\ref{ifree})
$$\langle {\tilde F}_\mu {\tilde F}_\nu \rangle \;=\;
D^{-1} (k^2) (\delta_{\mu\nu} - \frac{k_\mu k_\nu}{k^2})$$
$$\langle \theta \theta \rangle \;=\; \frac{D(k^2)}{k^2}$$
\be
\langle \ft \theta \rangle \;=\; \frac{k_\mu}{k^2}
\ee

We shall now consider some examples of application of the general recipe
to different models.

A simple example of an application would be to consider a model
defined by the parity-conserving functional
\be
I(\ft) \,=\, \int d^3 x \, \left[ \oh \ft \, D(-\partial^2) \,\ft \,+\,
\frac{g}{4 !} (\ft \ft)^2 \right]
\ee
where $D$ can be a complicated function of $\partial^2$, and $g$
is a coupling constant. The quartic term induces of course vertices
with four $\ft$ lines, and the theory is in that respect quite
simple. On should however be a bit careful due to the presence
of the Lagrange multiplier field $\theta$. Due to the quadratic
mixing it is better to regard $\ft$ and $\theta$ as two components
in some `internal space' of some field, and assign to the propagator
the corresponding matrix structure. This is useful when trying to find
the expression for Green's functions in terms of one-particle
irreducible ones.
The application of this procedure to the full $\ft$ propagator yields
\be
\langle \ft {\tilde F}_\nu \rangle \;=\;
\frac{1}{D(k^2) \,+\, \Pi^\perp (k^2)} \;
(\delta_{\mu\nu} - \frac{k_\mu k_\nu}{k^2})
\ee
where $\Pi^\perp$ is the transverse component of the irreducible
two-point
function for the field $\ft$
\be
\Pi_{\mu\nu} (k^2)\;=\; \Pi^\perp (k^2) (\delta_{\mu\nu} - \frac{k_\mu
k_\nu}{k^2})
\,+\, \Pi^\parallel (k^2)\frac{k_\mu k_\nu}{k^2} \;.
\ee
The mixed propagator $\langle \ft \theta \rangle$ does not renormalizes,
and
for the $\langle \theta \theta \rangle$  we obtain
\be
\langle \theta \theta \rangle \;=\; \frac{D(k^2) + \Pi^\parallel
(k^2)}{k^2} \;.
\ee
The one-loop correction to the effective action is easily computed
within this
scheme, and it is even quite straightforward to obtain calculate, in the
same
approximation, the effective action in the presence of an external
`monopole'
source $\rho$, introduced by modifying the Bianchi identity in the
following
way:
\be
\partial \cdot {\tilde F} \; \to \; \partial \cdot {\tilde F} \,-\, \rho
(x) \;.
\ee
An interesting example of an application
is the calculation of the static interaction energy between two (static)
monopoles,
defined as the part of the effective action depending on the distance
between
two localized static sources of strengths $\phi_1$ and $\phi_2$ located
at
${\vec x}_1$ and ${\vec x}_2$. The corresponding $\rho$ is defined by:
\be
\rho (x) \;=\; \phi_1 \delta (x_3) \delta ({\vec x} - {\vec x}_1) \,+\,
\phi_2 \delta (x_3) \delta ({\vec x} - {\vec x}_2)
\ee

The static energy density ${\cal E}({\vec x}_1 - {\vec x}_2)$ is given
by
$$
{\cal E}({\vec x}_1 - {\vec x}_2)\;=\;
\lim_{L \to \infty} \, \frac{1}{L^3} \, \int {\cal D} \ft \, {\cal D}
\theta \,
$$
\be
\frac{\exp \left\{ - I({\tilde F}) + i \int d^3 x \theta (\partial \cdot
{\tilde F}
- \rho) \right\}}{\exp \left\{ - I({\tilde F}) + i \int d^3 x \theta
(\partial \cdot
{\tilde F}) \right\}}
\ee
where $L$ is the length of the Euclidean box where the theory is
defined.

In the one-loop approximation, ${\cal E}$ becomes
\be
{\cal E}({\vec x}_1 - {\vec x}_2)\;=\; \phi_1 \phi_2 \; \gamma (r)
\ee
where $r = |{\vec r}| = {\vec x}_1 - {\vec x}_2$, and
\be
\gamma (r) \;=\; \int \frac{d^2 k}{(2 \pi)^2} \, e^{i {\vec k} \cdot
{\vec r}}
\frac{1}{(I^{-1})_{\mu\nu} (k)\,  k_\mu k_\nu}
\ee
with $I_{\mu\nu} = \frac{\delta^2 I}{\delta {\tilde F}_\mu (x_1)
\delta {\tilde F}_\nu (x_2)}$. This formula yields the interaction
potential
$\gamma$ as a complicated functional of the inputs of the effective
theory.

For the particular case of a static $\fmn$, and generalizing from the
quartic potential to a general one $V({\tilde F}^2)$, the form of
$\gamma$
can be further
simplified to
\be
\gamma (r) \;=\; \int \frac{d^2 k}{(2 \pi)^2} \, e^{i {\vec k} \cdot
{\vec r}}
\frac{1}{k^2 [ D(k^2) + 2 V'({\tilde F}^2)]} \;.
\ee

As another example, we note that
the situation, particular to $2+1$ dimensions, of $\ft$ being a one-form
field, allows we to construct action functionals $I$ depending only on
the `field strength' $W_{\mu\nu} = \partial_\mu {\tilde F}_\nu -
\partial_\nu {\tilde F}_\mu$. That is to say, one can consider models
where ${\tilde F}$ plays the role of a connection.
Any such functional $I$ will be
invariant under a new set of gauge transformations, defined as
\be
\ft \; \to \; \ft + \partial_\mu \omega \;.
\ee
This gauge invariance of $I$ allows us to regard now the constraint
$\partial \cdot {\tilde F}$ as a particular gauge fixing for this
symmetry, and thus to use a {\em different \/} gauge fixing without
affecting the physics. For example, one ends up with a generating
functional of the form:
\be
{\cal Z}(J_\mu) \;=\; \int \,{\cal D}\ft \,
e^{- I (\partial_\mu {\tilde F}_\nu - \partial_\nu {\tilde F}_\mu)
-\int d^3 x \frac{1}{2 \alpha} (\partial_\mu \ft )^2
+ \int d^3 x J_\mu {\tilde F}_\mu } \;.
\ee
when the family of covariant $\alpha$-gauges is used. Of course,
physical results should be independent of $\alpha$.
The physical meaning of this independence of physical results
on $\alpha$ would be at first sight surprising, since it means
that one can modify the Bianchi identity quite arbitrarily,
introducing monopoles into the play without altering the physics.
The reason is that, in the original variables,
this kind of model depends on $A_\mu$ only through the combination
$\partial_\mu F_{\mu\nu}$, namely
\be
S_{inv} (A) \;=\; {\cal F} (\partial_\mu F_{\mu\nu})
\ee
where ${\cal F}$.This automatically imposes the existence of second
derivatives for
$A_\mu$, forbidding the existence of monopoles.

\section*{Acknowledgments}
This work was supported by CONICET and Fundaci\' on
Antorchas.The author acknowledges Prof. R. C. Trinchero
for many useful comments.



\begin{references}
%
\bibitem{mig1}A. Migdal, Ann. Phys. {\bf 109}, 365 (1977).
\bibitem{mig2}Y. Makeenko and A. Migdal, Phys. Lett. B {\bf 88},
135 (1979); Nuc. Phys. B {\bf 188}, 269 (1981).
\bibitem{hal1}M. B. Halpern, Phys. Rev. D {\bf 16}, 1798 (1977).
\bibitem{hal2}M. B. Halpern, Phys. Rev. D {\bf 16}, 3515 (1977).
\bibitem{hal3}M. B. Halpern, Nucl. Phys. B {\bf 139}, 477 (1978).
\bibitem{hal4}M. B. Halpern, Phys. Rev. D {\bf 19}, 517 (1979).
\bibitem{ha1} P. E. Haagensen, K. Johnson, Nucl. Phys. B {\bf 439}, 
597 (1995). 
\bibitem{ha2}P. E. Haagensen, K. Johnson, C.S. Lam,
Nucl. Phys. B {\bf 477}, 273 (1996).
\bibitem{ait}I. J. R. Aitchison, C. D. Fosco and F. D. Mazzitelli,
Phys. Rev. D {\bf 54}, 4059 (1996).
\bibitem{mo1}T.R. Morris, Phys. Lett. B {\bf 357} (1995) 225.
\bibitem{mo2}T.R. Morris, Phys. Rev. D {\bf 53} (1996) 7250.
\bibitem{frad} E. Fradkin, {\em Field Theories of Condensed
Matter Systems}, Addison Wesley, Reading, 1991.
\end{references}
\end{document}